\documentclass{nature}
\usepackage{amsmath}
\usepackage{amssymb}
\usepackage{wasysym}
\usepackage[dvips]{graphicx}
\usepackage{cancel}
\usepackage{ulem}
\usepackage{xcolor}
\usepackage[latin1]{inputenc}
\usepackage{times}
\title{\textit{Confined linear carbon chains: a route to bulk carbyne}}

\author{Lei Shi$^1$, Philip Rohringer$^1$, Kazu Suenaga$^2$, Yoshiko Niimi$^2$, Jani Kotakoski$^1$, Jannik C. Meyer$^1$, Herwig Peterlik$^1$, Paola Ayala$^1$, \& Thomas Pichler$^{1*}$}

\begin{document}

\maketitle

\vspace{-0.5cm}

\begin{affiliations}
\item Faculty of Physics, University of Vienna, 1090 Wien, Austria
\vspace{-0.3cm}

\item National Institute of Advanced Industrial Science and
Technology (AIST), Nanotube Research Centre, 305-8565 Tsukuba,
Japan

\end{affiliations}

\vspace{-0.5cm}
\begin{abstract}
The extreme instability and strong chemical activity of carbyne,
the infinite \textbf{$sp^1$} hybridized carbon chain, are
responsible for its low possibility to survive at ambient
conditions. Therefore, much less has been possible to explore
about carbyne as compared to other novel carbon allotropes like
fullerenes, nanotubes and graphene. Although end-capping groups
can be used to stabilize a carbon chain, length limitation is
still a barrier for its actual production, and even more for
applications. Here, we report on a novel route for bulk
production of record long acethylenic linear carbon chains
protected by thin double-walled carbon nanotubes. A corresponding
extremely high Raman band is the first proof of a truly bulk yield
formation of very long arrangements, which is unambiguously
confirmed by transmission electron microscopy. Our
production establishes a way to exceptionally long stable
carbon chains, and an elegant forerunner towards the final goal of
a bulk production of essentially infinite carbyne.

\end{abstract}

\vspace{1.5cm}

\maketitle

Different kinds of allotropes can be formed from elemental carbon
due to its $sp^{n}$ hybridization\cite{Hirsch10NM}. Well known are
diamond and tetrahedral amorphous carbon as $sp^{3}$ hybridized
solids, whereas $sp^{2}$ hybridization is present in graphite,
fullerenes\cite{Kroto85N}, carbon nanotubes~(CNTs)\cite{Iijima91N}
and graphene\cite{Novoselov04S}. All of them have been intensively
investigated in the last decades. However, since carbyne was
proposed, as third allotrope as infinite chain with $sp^{1}$
hybridization in the sixties of the last century, its research
attracted great interest but also significant
controversy\cite{Kasatochkin67DANS,Goresy68S,Whittaker69S,Sladkov69P}.
For several years the identification of carbyne remained
questionable\cite{Smith81S,Smith82S} until the experimental
accomplishment to unambiguously produce polyynes: short $sp^{1}$
hybridized linear carbon arrangements with end-capping groups. The
longest reported polyynes so far consist of 44 contiguous carbon
atoms with alternating single and triple
bonds\cite{Chalifoux10NC}, but the bulk synthesis of carbyne or
very long linear carbon chains (LLCCs) continues to face stumbling
blocks. Very little is known about the $sp^{1}$ hybridization in
carbyne because of its extreme instability in ambient conditions.
In fact, a longstanding study postulated the impossibility to
prepare this truly 1D material\cite{Bayer85BdDCG}. Although the
same had been foreseen for graphene on account of its
thermodynamic instability, its rise initiated by a feasible
synthesis route and the later understanding of its unique
properties\cite{Novoselov04S} have inspired revisiting carbyne
from the theoretical and experimental points of view. Carbyne's
theoretically anticipated higher strength, elastic modulus and
stiffness than any known material, including diamond, carbon
nanotubes and graphene allows envisioning new
composite materials\cite{Liu13AN}. Another example of essentially unexplored
applications would be having the utmost limit in channel-width for
field-effect transistors given by a one-carbon-atom
thickness\cite{Baughman06S}. The experimental research in this
field has been heavily focused on the synthesis of linear carbon
chains by different methods, among which, heavy end-capping groups
have commonly been used to stabilize
them~\cite{Eisler05JotACS,Chalifoux10NC}. At the same time,
single-(SW), double-(DW) and multi-walled (MW) CNTs have
increasingly been used as confining nanoreactors where novel
one-dimensional materials can be produced such as: short
polyynes\cite{Nishide06CPL}, metal nanowires\cite{Kitaura08NL} and
ultra-narrow graphene\cite{Chuvilin11NM}. MWCNTs have been
reported as hosts of carbon nanowires with about 100 atoms using a
direct arc-discharge process\cite{Zhao03PRL}. Also, the growth of
short carbon chains from high-temperature treatments of
buckypapers has been accomplished in DWCNTs\cite{Endo06S}.
These have also hosted the production of carbon nanowires upon
fusion of molecules like C$_{10}$H$_2$ or
adamantane\cite{Zhao11JoPCC,Zhang12AN}.

In this study, we have sought to use the full potential of DWCNTs
bringing into line their function as nanoreactors and as a method
to protect LLCCs from being decomposed. They are not only nicely
suitable for a nanoreactor function, but their narrow diameter can
also promote the encapsulation of LLCCs\cite{Zhao03PRL,Shi11NR}.
Our procedure uses very high temperatures and high vacuum (HV)
conditions, which allows the bulk realization of the longest
existing LLCCs consisting of more than 10000 contiguous acetylenic
carbons inside thin DWCNTs with ideal diameters. The following results involving the high recurrence of
LLCC inside DWCNTs (LLCC@DWCNTs) substantiate a truly feasible
path towards the formation of carbyne at bulk scale.

\begin{figure*}
\begin{center}
\includegraphics[width=14cm]{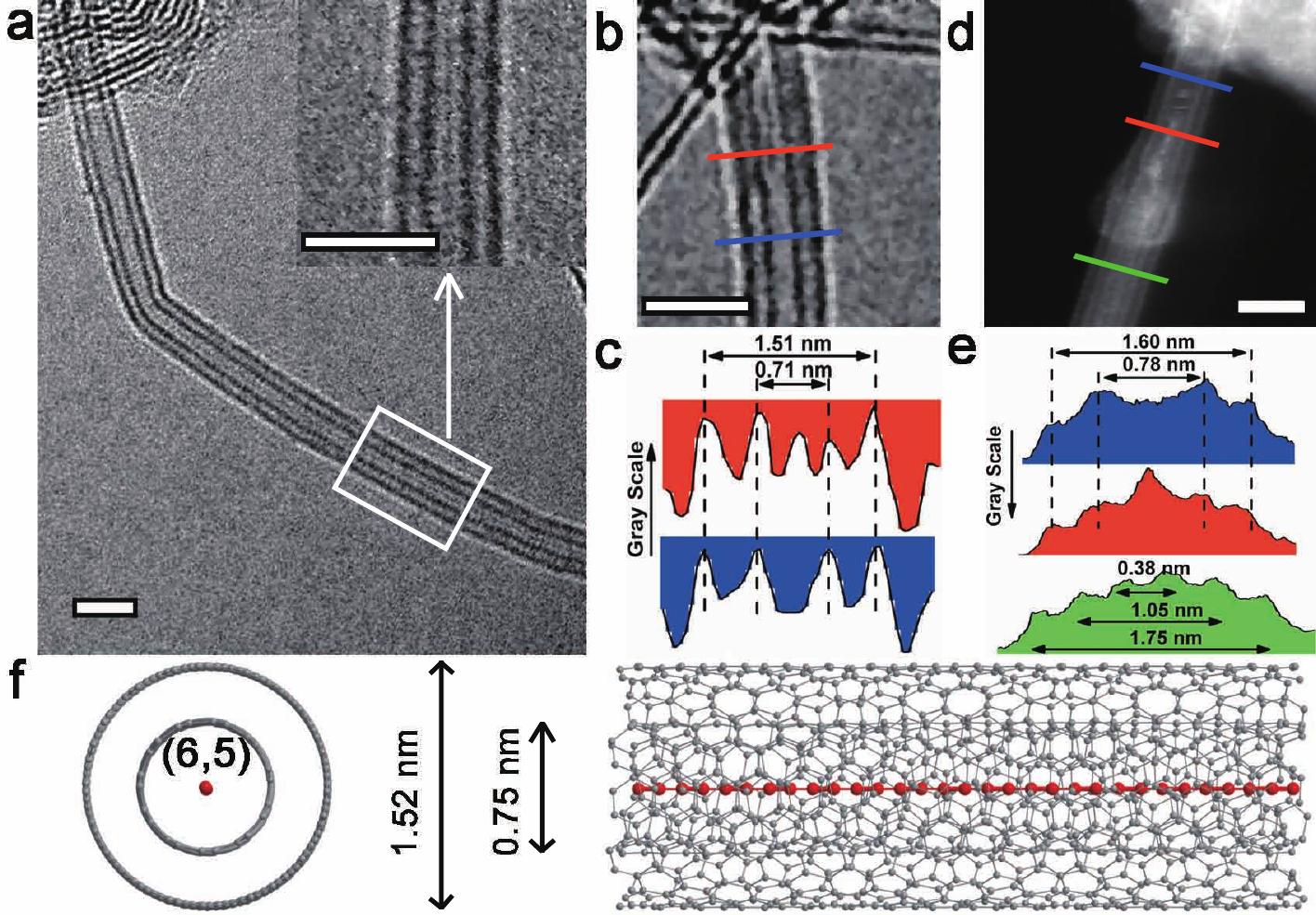}
\end{center}
\vspace{-0.7cm} \caption{\textsf{\textbf{Direct observation of
LLCC@DWCNT.} ~\textbf{a}, HRTEM image of a LLCC@DWCNT with
bending. The LLCC inside a DWCNT is longer than 26 nm, which means
that it consists of more than 200 continuous carbon atoms. Inset:
A close view of the image in the marked box. ~\textbf{b}, Another
part of DWCNT with half filling of LLCC. ~\textbf{c}, The line
profiles at positions along the blue and red lines shown in b,
represent empty DWCNT and LLCC@DWCNT, respectively. ~\textbf{d},
The STEM image of a LLCC@DWCNT. ~\textbf{e}, The line profiles at
positions along the blue, red and green lines shown in (d),
represent empty DWCNT, LLCC@DWCNT and a thin most-inner
tube@DWCNT, respectively. ~\textbf{f}, Molecular model of
LLCC@DWCNT with (6,5) inner tube. Scale bar: 2 nm.}} \label{Fig1}
\end{figure*}

Our direct observations on the LLCCs by aberration corrected
high-resolution (HR) and scanning (S) transmission electron
microscopy (TEM) are summarized in Figure~\ref{Fig1}. Our
DWCNT--hosting--material is characterized by a high yield of very
small diameter inner tubes ($\sim$0.65 to 0.9~nm), where (6,4),
(6,5) and (8,3) are the most common inner chiralities. The
HRTEM micrograph in Fig.~\ref{Fig1}a shows a long bent DWCNT
encapsulating a carbon chain. Taking into account that the
single and triple bond-lengths are 0.1229 and 0.1329 nm exhibiting a calculated lattice constant of d=0.2558 nm, more than 200 contiguous carbon atoms
are contained within the $\sim$26~nm. This perfectly matches our complementary bulk scale small angle x-ray diffraction results revealing a lattice constant of d=0.252 nm (see supplementary Figure S18). To the
best of our knowledge, the LLCC has by far the directly in TEM
observed length record and extremely built-up production: many
tubes are filled at least partially as shown in Fig.~\ref{Fig1}b.
These TEM studies also suggest that the LLCCs can easily withstand
the bending of the DWCNT, because breaking a carbon-carbon bond needs much
more energy than the strain energy introduced by
bending\cite{Hu11JoPCC}. 

The LLCCs actually always move during the HRTEM observations and
therefore appear blurry compared to the DWCNT's walls. This
entails a profile analysis of the empty and filled parts of the
DWCNT. In Fig.~\ref{Fig1}c In Fig. 1c the additional peak shows the difference when the LLCC exists inside DWCNT. Complementary STEM studies on a different tube shown in
Fig.~\ref{Fig1}d and the corresponding scanning blue and red
profiles (Fig.~\ref{Fig1}e) confirm the observation by HRTEM. The green profile corresponds to
a thicker tube merging a thinner tube. In this case a third inner tube
was formed instead of a LLCC@DWCNT, which was consistently
observed in all samples. This allows us generalizing that the HV
high temperature nanochemical reactions form LLCCs only for DWCNT
with outer diameters below $\sim$ 1.75nm. For slightly thicker
DWCNT, triple-walled CNT are formed and significantly larger ones
host twisted carbon-clusters (see supplementary information). This
establishes the diameter as the key factor in the synthesis of
LLCCs. To better visualize this, we have included a
model of a LLCC inside a DWCNT  with a (6,5) inner tube of 0.75
nm, which is close to the ones observed in HRTEM. Although the
HRTEM and STEM are extremely helpful, the fine structure of the
LLCC remains elusive due to the movement of the LLCC inside the
DWCNT.

\begin{figure*}
\begin{center}
\includegraphics[width=14cm]{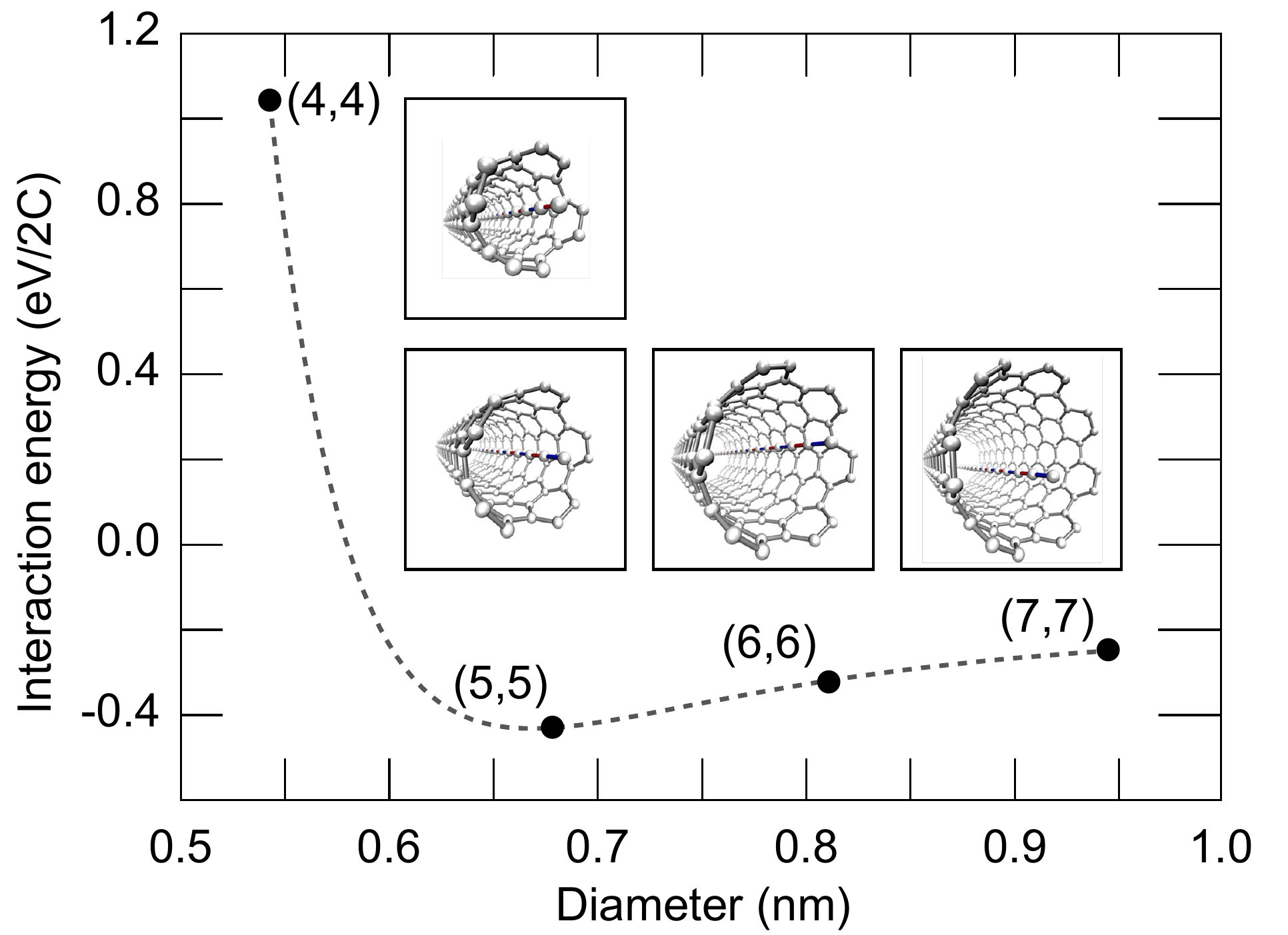}
\end{center}
\caption{\textsf{\textbf{Calculation of the optimum diameter for
LLCC encapsulated in CNT.} DFT calculations about the interaction
energies between the LLCC and different chirality tubes: (4,4),
(5,5), (6,6) and (7,7). The inset: illustrations of different
CC@CNT structures. Notice that the distance between the LLCC and the wall of
the CNT remains similar for all host CNT. For the energy minimum, this distance is close to the radius of the CNT.}} \label{Fig2}
\end{figure*}

To further understand the effect of confinement on the growth of
LLCCs, we performed density functional theory calculations with
the Vienna ab--initio simulation package
(VASP)\cite{Kresse96CMS,Kresse96PRB} using projector augmented
wave potentials\cite{Bloechl94PRB}. The stable chain configuration
turned out to be exactly straight, and to contain alternating
single and triple bonds with bond lengths of 1.329 \AA \, and
1.229 \AA, respectively. We limited our calculations to four small
armchair nanotubes, namely, (4,4), (5,5), (6,6) and (7,7) with one
repeating unit in the length direction. The interaction energies
between the carbon chains and the different nanotubes are plotted
in Fig.~\ref{Fig2}, and the inset shows the illustrations of the
LLCC inside different CNTs. The result shows that
the optimum distance between adjacent carbons and the CNT is
0.3378 nm, which is very close to the optimum inter-layer distance
between graphite layers ($\sim$0.32 nm), as described in the
DFT-D2 method\cite{Graziano12JoPCM}. The LLCC@(5,5) exhibit the
lowest interaction energy, which means that a (5,5) nanotube with
a diameter of 0.69 nm is the optimum among all four different
proposed CNTs for the LLCC--growth. Assuming that the ratio
between the carbyne and CNT distance and the inter-layer distance
were the same in the calculations and in the experiments (i. e.
scaling theoretical values to experimental ones), this would lead
to a predicted optimum diameter of
2$\times$0.3378$\div$0.32$\times$0.335 nm = $\sim$0.71 nm, which
is in a very good agreement with the microscopy outcome.

\begin{figure*}
\begin{center}
\includegraphics[width=12cm]{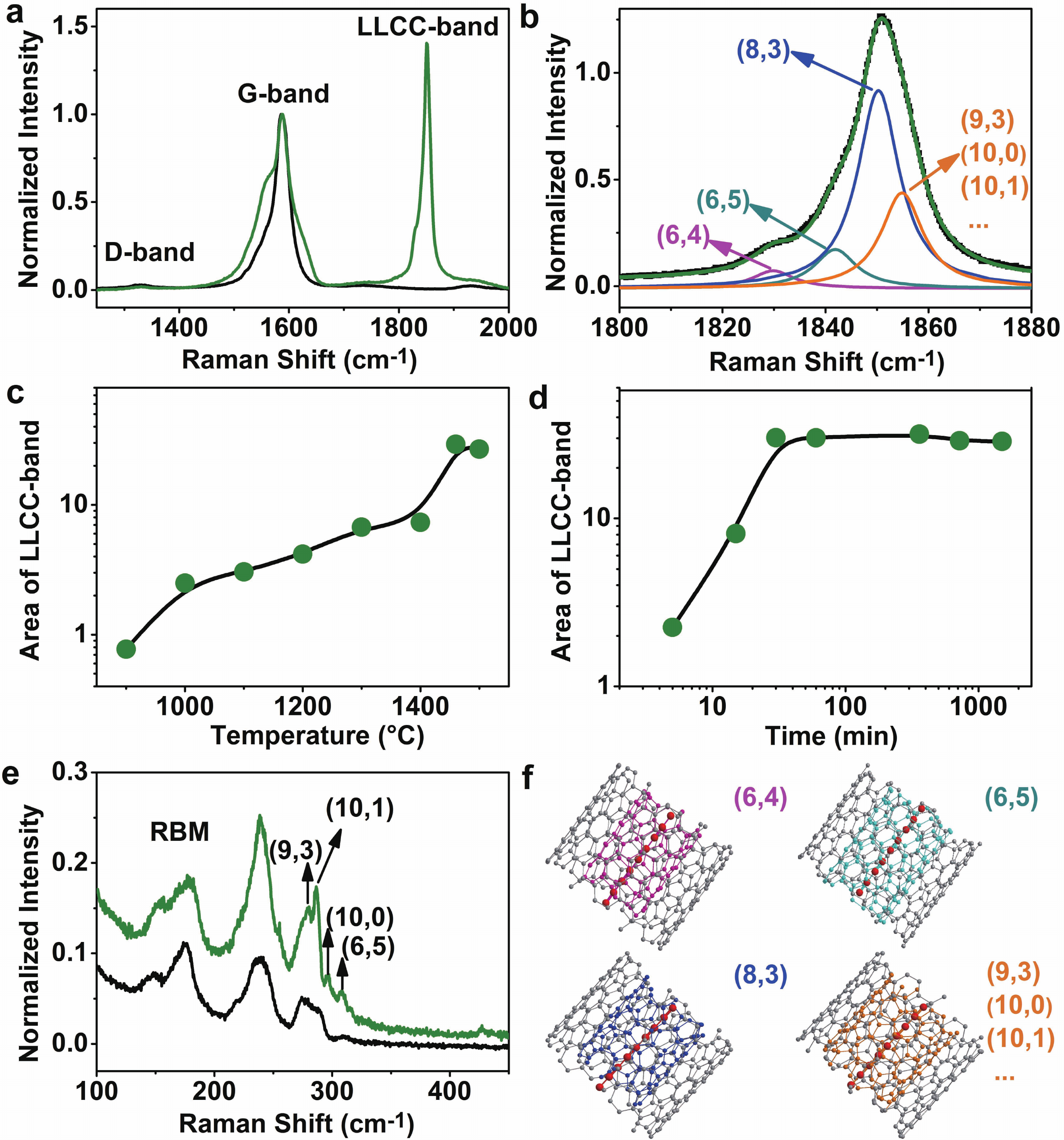}
\end{center}
\caption{\textsf{\textbf{Raman spectra of LLCC@DWCNTs.} These
measurements were done using a 568.2~nm excitation wavelength.
~\textbf{a}, D-band, G-band and LLCC-bands of a sample annealed at
1460 $\,^{\circ}\mathrm{C}$ as compared to a pristine DWCNT. ~\textbf{b}, LLCC-band line shape analysis including four components. ~\textbf{c}, The area of the
LLCC-band as a function of annealing temperature for 30 min
annealing time. ~\textbf{d}, The area of the LLCC-band as a
function of annealing time at 1460 $\,^{\circ}\mathrm{C}$.
~\textbf{e}, RBM region for a sample annealed
at 1460 $\,^{\circ}\mathrm{C}$. ~\textbf{f}, Schemes of our most
abundant inner-tube-chiralites and chain
configurations.}}\label{Fig3}
\end{figure*}

\begin{figure*}
\begin{center}
\includegraphics[width=16cm]{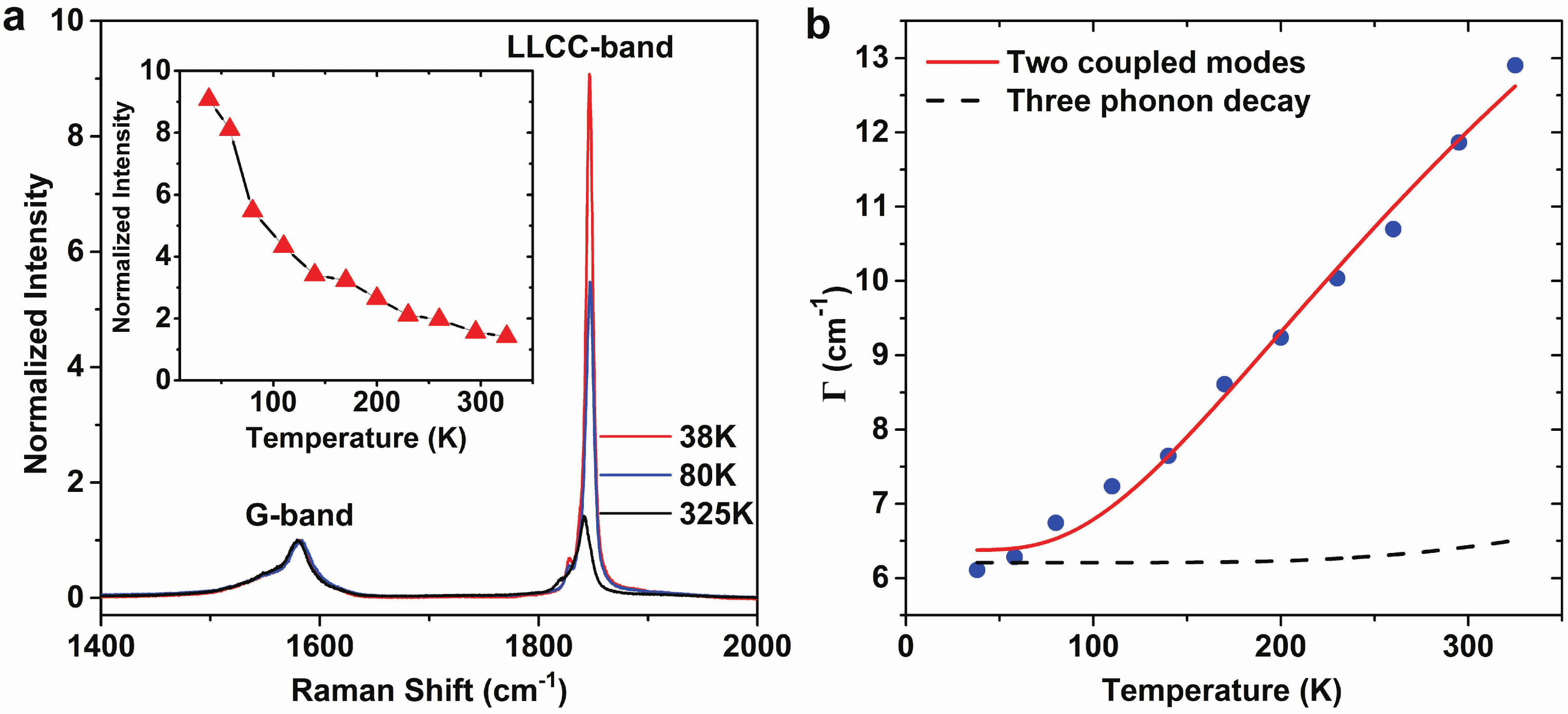}
\end{center}
\caption{\textsf{\textbf{The temperature dependence of LLCC-band.}
~\textbf{a}, Raman spectra of LLCC@DWCNTs measured at different
temperatures. The spectra are are normalized to the G-band response. The inset shows the LLCC-band intensity in comparison to the G-line.
~\textbf{b}, FWHM of the individual components in LLCC-band as a
function of temperature. The solid line represents a model of a temperature dependent coupling two coupled modes, the dashed curve represents a fit to a model of a three-phonon decay. }}
\label{Fig4}
\end{figure*}

To verify the optimal DWCNT diameter and growth conditions for
bulk yield synthesis of LLCCs, resonance Raman spectroscopy was used. The
LLCCs exhibit a Raman active
mode\cite{Zhao11JoPCC,Zhao03PRL,Shi11NR}, named hereafter as
LLCC-band. Its Raman frequency corresponding is closely associated
to the length of the LLCCs and its intensity to the overall yield.
Polyyne molecules with less than 20 carbon atoms have vibrational
features between 1900-2300 cm$^{-1}$,\cite{Agarwal13JoRS} whereas in LLCCs consisting
of about 100 atoms inside MWCNTs have reported a response at
$\sim$1825-1850 cm$^{-1}$.\cite{Zhao03PRL} Our multi-frequency studies using a dye
laser, confirmed a broad resonance window of the LLCC-band peaking
at about 568 nm for LLCC with above 6 different length of acetylenic chains.(see
supplementary information). As shown in Fig.~\ref{Fig3}, inspecting the characteristic D
and G bands, corresponding to the defect induced and the graphitic
mode of pristine DWCNTs and DWCNTs with LLCCs grown at
1460$\,^{\circ}\mathrm{C}$ (Fig.~\ref{Fig3}a), the D/G intensity
ratio remains practically unchanged. This highlights that
the DWCNTs are not damaged during annealing. Moreover, after
annealing a very intense LLCC band with fine structure at
$\sim$1850 cm$^{-1}$ is observed. A detailed line shape analysis
is shown in Fig.~\ref{Fig3}b. The Raman response of the LLCC-band
was always found between 1780 and 1880 cm$^{-1}$ concomitant with
an intensity several times larger than in any previous works on
polyynes reaching a record of an even nine times bigger intensity
as compared to the DWCNT G-band response at low temperature (see
Fig.~\ref{Fig4}a below). This confirms that the chains are always
very long and that their production is feasible in bulk
quantities. However, the actual yield of the grown LLCC strongly
depends on the growth time and growth temperature in HV.
Fig.~\ref{Fig3}c summarizes the changes in the area of the
LLCC-band as a function of annealing temperature. Our new HV method allows to
synthesize a small quantity of LLCCs even at a temperature as low
as 900$\,^{\circ}\mathrm{C}$. This is consistant with reports on
fusion of linear polyynes or adamantane molecules inside DWCNTs at
similar temperatures, but those linear structures grow in
relatively low yields depending on the filling ratio
\cite{Zhao11JoPCC,Zhang12AN}. The highest area of the LLCC-band
indicats that 1460$\,^{\circ}\mathrm{C}$ is the optimal temperature to form LLCCs. Further
increasing the annealing temperature leads to a subsequent
decomposition of the LLCC@DWCNT structure. Aiming at
understanding the growth kinetics of the LLCCs, the pristine
DWCNTs were also annealed in HV at the optimal temperature of
1460$\,^{\circ}\mathrm{C}$ for different time. As observed in
Fig.~\ref{Fig3}d, after five minutes annealing, the LLCC-band
appears and its intensity increasing for annealing times up to 30
min and saturates afterwards. This suggests that growth of the
LLCCs is finished. However, a line shape analysis of the LLCCs
(similar to Fig. 3b) revealed that for longer annealing time the
relative intensity of the peaks at lower frequency get
progressively stronger, indicating the continuous growth of longer
chains. The decisive aspects for the LLCC production on top of our
studies are: the small-diameter inner tube of our DWCNTs and the
annealing in HV, which allows the free carbon atoms having a longer
mean free path. It is also worth mentioning that SWCNTs below
$\sim$0.7 nm diameter are not stable at high temperature and
therefore only very few LLCCs can be grown inside them (see
supplementary information).

\par

A remaining important question is how the LLCC structure is
confined inside the tubes. This missing puzzle piece can be
retrieved from an analysis of the radial breathing modes (RBM) of
DWCNTs depicted in Fig.~\ref{Fig3}e. These DWCNTs can be indexed
by their chiralities using the Kataura plot and the established
inverse proportionality of the tube diameter and the RBM frequency
including a correction term for the intertube
distance\cite{Kuzmany01EPJB,Araujo08PRB}. The indexed RBM peaks
corresponding to the thin inner tubes are split and enhanced. This
is a first indication of an interaction with the LLCCs. Turning to a detailed analysis
of the yield and length distribution of the LLCCs, a tunable dye laser, e.g. with 588.8
and 606.0 nm excitations was used to reveal
the full pattern of the LLCC-band (see supplementary information). In general, the LLCC-band position is
correlated to the length of the LLCC following the bond length
alternation model with a hybrid force field\cite{Yang07JoPCA}.
However, this model is limited regarding an absolute length
determination and handles environmental interactions empirically.
Considering the interaction between the LLCC and the inner tubes,
for thinner host tubes the interaction is larger and therefore a
lower frequency is observed. This enables us to tentatively assign
the LLCCs fine structure in the line shape analysis of
Fig.~\ref{Fig3}b to different chiralities of inner host tubes.
Considering the total energy of the LLCC@DWCNTs, (6,4) tubes are
more prone to host longer LLCCs than (6,5). At the same time, a
(6,4) tube has larger interaction with its encapsulated LLCC than
a (6,5) one. Further exploring the fine structure LLCC band allows
estimating the growth yield of different LLCC@DWCNTs via the
different relative intensities of the LLCC-bands. This calibration
allows easily comparing the growth yields of LLCCs among different
samples. This analysis relies on a weak van der Waals coupling
between the inner tubes and the LLCCs, which has to be verified.

In order to do this, further information on the LLCCs was then
sought analysing the temperature dependent Raman response. The
intensity and area of LLCC-band are much higher at low
temperatures as shown in Fig.~\ref{Fig4}a. The data recorded from
38 to 325 K are plotted in the inset and show an intensity of
156\% of the G-line intensity at room temperature, which increases
to 908\% of the G-line intensity at 38 K. Concomitantly, it was
observed that the full width at half maximum (FWHM, $\Gamma$) gets
noticeably narrowed down (Fig.~\ref{Fig4}b). Normally, the
temperature dependence of the FWHM can be analyzed by a classical
model of anharmonic decay of optical
phonons\cite{Cowley65JDP,Klemens66PR,Balkanski83PRB} or a model of
a temperature dependent coupling of two modes. As shown in
Fig.~\ref{Fig4}b, the anharmonic decay does not fit our
experimental results at all. Hence we describe our results by
the second model, i.e. a coupling of the LLCCs mode to the RBM of the
inner tubes. The temperature dependence of the FWHM follows the
equation with activation energy E$_a$\cite{Kuzmany1998}.
 \begin{equation*}
 \ \Gamma(T) = \Gamma(0) ({1 + {e^{-E_a/k_{B}T}}})
 \end{equation*}
As plotted as solid line in Fig. 4b, we
find a perfect match for an activation energy E$_a$ of 33.85 meV.
This is a clear proof of the coupling of the LLCCs to the inner
tubes of the DWCNTs and highlights the importance of the confined
nanospace with the right diameter introduced above. This
activation energy is also consistent with a weak van der Waals
interaction between the tubes and the LLCCs.

\begin{figure*}
\begin{center}
\includegraphics[width=14cm]{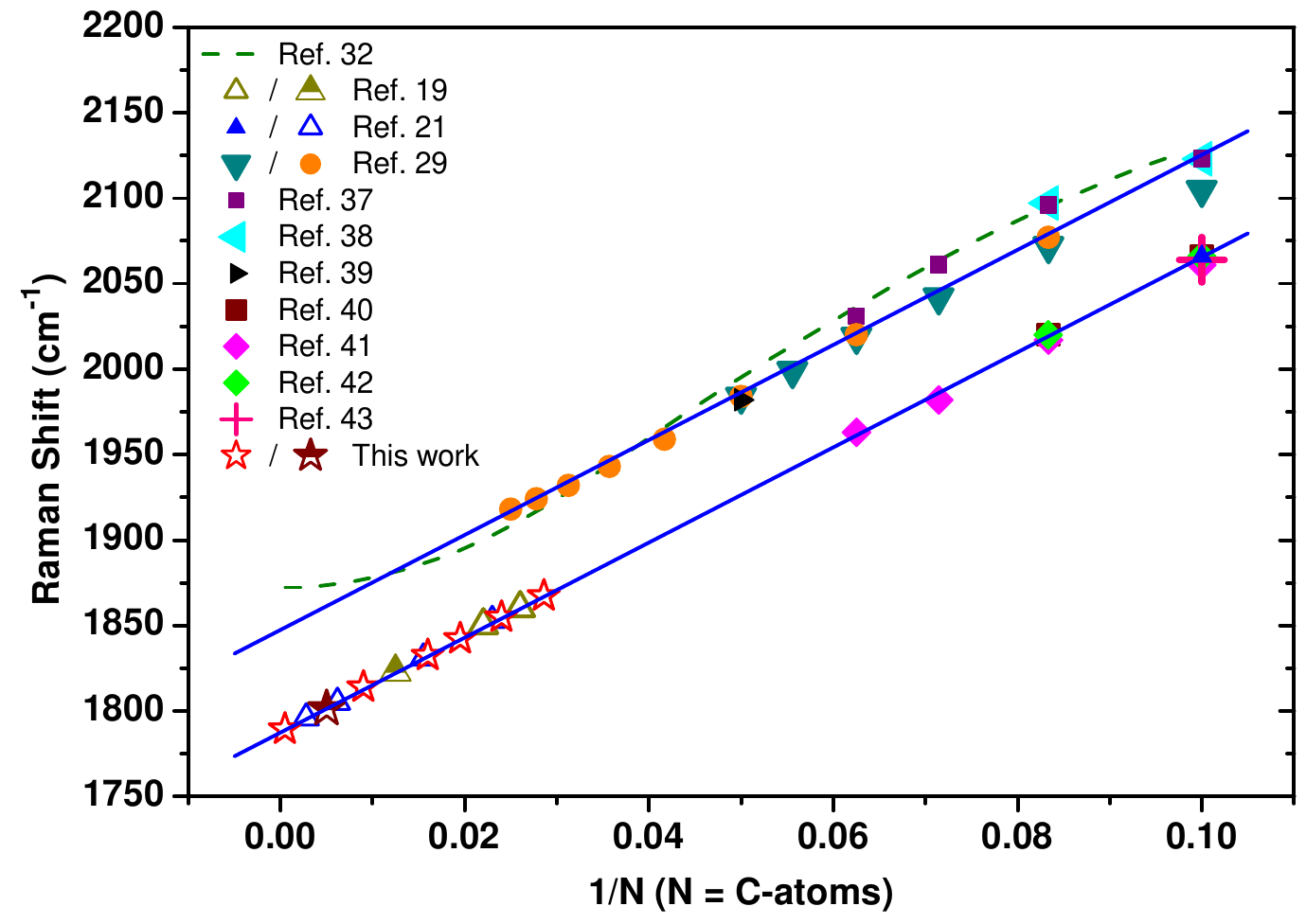}
\end{center}
\caption{\textsf{\textbf{Comparison of the inverse lengths of
linear carbon chains with the Raman response}. Raman response of
polyynes as function of inverse length given by the number of
carbon atoms. The solid lines correspond to a linear fit. The top
line includes the data related to chains that are saturated or
suspended in a fluid medium \cite{Tabata06C,Wakabayashi07CPL,Agarwal13JoRS,Gibtner02CEJ}. The lower line contains
studies all embracing nanotubes taking into account an additional
downshift of 60 cm$^{-1}$ due to the interaction with the inner
tubes. The solid symbols are experimental results with confirmed
length \cite{Malard07PRB,Wakabayashi09EPJD,Nishide07JoPCC,Moura11PRB,Zhao11JoPCC} and the empty symbols correspond to the LLCC@DWCNTs
reported in this work and for LLCC@D/MWCNT \cite{Zhao11JoPCC,Zhao03PRL}. The partly
solid symbols are experimental results with length estimated by
TEM. The dashed line corresponds to a theoretical model using the
bond length alternation from Yang. et al.\cite{Yang07JoPCA}. }} \label{Fig5}
\end{figure*}

In keeping with these results we are able to put the analysis one
step further and unambiguously correlate the length of the LLCCs
with the Raman position. In Fig.~\ref{Fig5} we compare the Raman
positions of different linear carbon chains reported in the
literature with the theoretical model from Yang et
al.\cite{Yang07JoPCA} (dashed line) as a function of the inverse
carbon atom number. Although there is a qualitative agreement, the
limitations of the model are obvious and further precise
calculations are needed. However, strikingly we find a very good
linear correlation for the line positions of the experimentally
proven inverse chain length. This is highlighted by the solid
symbols. This linear correlation also counts for LCCs in different
environments. LCC inside the same environment follow the same
curve, but the LCC in liquid environment is about 60 cm$^{-1}$
higher in frequency as the ones inside CNT environment.

This allows to assign the length of the long LLCCs by Raman
spectroscopy. From our TEM results we know to have DWCNT
containing the at least 26 nm LLCC.  In order to correlate it with
the Raman response we did the most conservative approximation as
lowest limit for the chain length and took an extension of the
linear fit and assigned the Raman peak at about 1805 cm$^{-1}$ to the
26 nm chain. We are aware, that for infinitely long chains such a linear approximation is
not accurate anymore because of two competing effects. The bond length alternation will saturate,
yielding a blue-shift whereas a charge transfer, and hybridisation
with the inner tubes will yield a red-shift in the Raman response.
Nevertheless, the linear relation is a good first approximation and
allows us to assign all other observed Raman signals from our line
shape analysis and from the literature to an estimated chain length
(open symbols in Fig. 5). This tells us that our highest Raman
response corresponds to a chain of at least 40 continuous carbon
atoms and we have at lower yield several tubes containing
extremely long linear carbon chains of more than 10000 atoms. More detailed calculations
will allow to derive the exact anchor point for the chain length.
We prove that even with the most conservative estimation we get
bulk yield of LLCCs which are at least two order of magnitude
longer than those synthesized by an organic chemistry
route\cite{Chalifoux10NC}.

\par
While the exact LLCC@DWCNT formation mechanism via HV high
temperature nanochemical reactions remains unknown, our in-depth
inspection of the material allows us to suggest a model based on
the confinement inside the circular nanospace of tubes.
Considering the lowest energy principle, carbon atoms can move
inside the nanotubes from the open caps or through the tube walls
during the HV annealing process. The tubes with diameter
of $\sim$0.71 nm are the best suited to grow LLCCs as slightly
larger tubes tend to form spiral chains or if the tubes are too
thick in diameter the tubes are no longer working as nanoreactors
and the LLCCs are disintegrating. With our narrow diameter
distribution of inner tubes and high yield of small diameter
DWCNTs we are able to achieve a bulk production of LLCCs. If uniformed CNTs or the tubes with same
chirality are used as the nanoreactors, and providing additional
carbon supply, it is a route to achieve the growth of bulk carbyne
which might be further extracted from the nanotubes and stabilized
in different environments.

Summarizing, we have developed a new route for the first truly bulk
production of LLCCs, i.e. carbyne, confined inside DWCNTs. From our Raman
and TEM studies with contemporary ab-initio calculations we prove that stability, growth conditions as well as the
length and yield of the LLCCs crucially depend on the size
confinement. By analyzing the growth temperature dependence, we unambiguously revealed that
there is a coupling between the LLCCs and the inner tubes as
strongly supported by a model of coupled Raman modes. This
allows us to directly correlate the LLCCs length to the
diameter of the inner tubes and determine the optimal diameter 
for the high yield growth of carbyne (about 0.71 nm). 
Our approach of HV high temperature annealing has a great advantage over
filling SWCNT/DWCNT by polyynes and converting them to carbyne because of the superior yield achieving
real bulk samples, This is proven by an extremely huge Raman mode with
the intensity of more than 900
$\%$ of the G-band at 38 K in-keeping with a bulk lattice constant of the confined LLCCs chains of d=0.252 nm.  
Our longest LLCCs can be correlated to a chain length of more than 10000 carbon atoms, which can be seen as the closest realization of carbyne sofar.   \par
Inside the DWCNTs the LLCCs are very stable and with high yield, which is of great
important for further applications. Theoretical studies have shown
that after inserting LLCCs inside the CNTs, the hybrid system
would show metallic character, due to charge transfer from CNTs to
the LLCCs, although both of the CNTs and LLCCs perform
semiconductor properties themselves\cite{Rusznyak05PRB,Tapia10C}.
Therefore, we suggest that it is possible to control the
electronic properties (tunable band gap) of the hybrid system by
different filling yield of LLCCs. Furthermore, quantum spin
transport in LLCC can also be a promising application according to
the theoretical predictions\cite{Zanolli10AN}. Consequently, as
true 1D nanocarbon, this novel LLCC@DWCNT system would be, beside
the basic interest in chemistry, also a fascinating candidate for
the next generation of nanoelectronic devices. As a last point the LLCCs might also be further extracted from the DWCNTs
and stabilized in liquid environment later on.

\vspace{1cm}

\indent{\large\bf METHODS}

\subsection{DWCNT synthesis.}
 DWCNTs were synthesized by high vacuum alcohol chemical vapor
deposition. The catalysts were faint yellow powders obtained from
the mixture of ammonium iron citrate (Sigma-Aldrich, 3 wt.$\%$)
and MgO (Sigma-Aldrich, 97 wt.$\%$), which were suspended in
ethanol, sonicated for 1 hour and calcined at
70~$\,^{\circ}\mathrm{C}$ for 24 hours \cite{Endo05N}. The powder
was then placed in an alumina crucible inside a quartz tube
(diameter: 5 cm) for synthesis. The vacuum of the system was
always better than 10$^{-7}$ mbar, when 875 $\,^{\circ}\mathrm{C}$
(optimized growth temperature) was reached. An ethanol flow was
introduced into the quartz tube for 30 min as carbon source, and
the pressure was kept at $\sim$70 mbar during synthesis.
Typically, 0.5 g of catalyst powder yielded tens of milligrams of
DWCNTs after a subsequent purification processes. The first
nanotube purification step was done with immersion in HCl
($\sim$37 wt.$\%$). The second consisted in heat treatment in air
at 400~$\,^{\circ}\mathrm{C}$ for 30 min to remove the amorphous
carbon. Black powder was obtained and it was then immersed once
again in a HCl solution to remove any residual compound from the
catalysts. Subsequently, a last annealing treatment in air was
done at 500 $\,^{\circ}\mathrm{C}$ for 2 hours to remove
completely the amorphous carbon and chemically active
SWNTs\cite{Endo05N}. Finally, a thin DWCNT buckypaper was prepared
by rinsing, filtering and drying.

\subsection{Synthesis of LLCC@DWCNTs.}
The buckypapers were annealed under high vacuum conditions. Once
the optimal synthesis temperature was determined, the pressure
was lower than 8$\times$10$^{-7}$ mbar at various temperatures
(900-1500 $\,^{\circ}\mathrm{C}$) for 30 min. To understand the
growth process and to test the stability of the LLCCs, the DWCNT
buckypapers were also annealed at 1460 $\,^{\circ}\mathrm{C}$ for
different time intervals (5-1500 min).

\subsection{Sample characterizations.}
Scanning electron microscope (SEM, Zeiss Supra 55 VP) was used to
observe the morphology and abundance of CNTs in the as-grown
sample. The versatility of resonance Raman spectroscopy was used to
study the CNTs and LLCCs. The as-grown, purified DWCNT and HiPco
samples were measured with 568.2 (Kr$^{+}$ laser), 587.2 and 604.3
nm (dye laser, Rhodamine 6G) excitations in ambient conditions
using a triple monochromator Raman spectrometer (Dilor XY with a
liquid nitrogen cooled CCD detector, without objective lens, 2
mW). All the annealed samples were measured with excitation
wavelength of 568.2 nm, because this laser energy is close to the
resonance energy of the LLCCs. The slit width was set at 200
$\mu$m and the spectral resolution was about 2 cm$^{-1}$. Low
temperature Raman spectra was obtained with
excitation wavelength of 568.2 nm combined with a cooling system (Modle 22C Cryodyne Cryocooler with a temperature controller: Model DRC-91C). For ease of comparison, all the
spectra were normalized according to the intensity of the G-band.
Aberration-corrected HRTEM observations operated at 120 kV (2010F,
JEOL) were carried out to confirm the purity of the samples and to
directly proof the structure of LLCC@DWCNT. STEM (Nion UltraSTEM 100) with a medium annular dark field detector
operated at 60 kV was also used to confirm the the structure of LLCC@DWCNT.

\subsection{DFT calculations.}
To improve our understanding on the optimal nanotube diameter
for the growth of LLCCs, we performed density functional theory calculations
with the VASP code\cite{Kresse96CMS,Kresse96PRB} using projector
augmented wave potentials\cite{Bloechl94PRB}. In order to properly
describe the Peierls distortion, we utilized in our simulations
the HSE06 hybrid functional for exchange and
correlation\cite{Paier6JOCP}. The kinetic energy cutoff was set to
500 eV. All of the structures were placed in a 1.5 nm $\times$ 1.5
nm simulation cell in the xy-plane with the length in the
z-direction defined by the equilibrium length for a 2-atomic unit
cell of a carbyne chain. The scheme by Monkhorst and Pack was used
to generate a $\Gamma$-point centered 1$\times$1$\times$14 k-point
mesh for the integration in the reciprocal
space\cite{Monkhorst76PRB}. Van der Waals interaction was taken
into account via the DFT-D2 method of Grimme\cite{Grimme06JoCC}.

\vspace{1cm}

Supplementary Information accompanies the paper at
www.nature.com/naturematerials

\indent {\large \bf Acknowledgements}

This work was supported by the Austrian Science Funds (FWF). L.S.
thanks the scholarship supported by the China Scholarship Council. K.S. and Y.N.
acknowledge the JST research acceleration programme.
J. K. acknowledges FWF for funding through project M 1481-N20, as
well as Vienna Scientific Cluster for computational time, and he
also thanks Georg Kresse and Martijn Marsman for their help in DFT
calculations. We acknowledge Hans Kuzmany for his constructive
discussion about the low-temperature Raman spectra of
LLCCs@DWCNTs. We thank Stephan Puchegger from the Faculty Center for Nanostructure Research for the support with SEM imaging.

\vspace{1cm}

\indent {\large \bf Author Contributions}

All authors contributed to this work. L.S. and T.P. designed and supervised the
experiments. L.S. prepared the samples. L.S. and P.R. did
characterization with Raman. K.S. and Y.N. performed
HRTEM characterization. J.K. and J. M. did the STEM measurement.
J.K. did the DFT calculations. H.P. did the XRD measurement. L.S.,
P.A. and T.P. analyzed data and wrote the manuscript and the
supplementary information. All authors discussed the results and
commented on the manuscript at all stages. \vspace{1cm}

\indent {\large \bf Competing Financial Interests} The authors
declare that they have no competing financial interests associated
to the publication of this manuscript.

\indent {\large \bf Corresponding Author} thomas.pichler@univie.ac.at

\indent

\clearpage

\indent {\large \bf Figure Legends}

\textsf{\textbf{Figure1: Direct observation of
LLCC@DWCNT.} ~\textbf{a}, HRTEM image of a LLCC@DWCNT with
bending. The LLCC inside a DWCNT is longer than 26 nm, which means
that it consists of more than 200 continuous carbon atoms. Inset:
A close view of the image in the marked box. ~\textbf{b}, Another
part of DWCNT with half filling of LLCC. ~\textbf{c}, The line
profiles at positions along the blue and red lines shown in b,
represent empty DWCNT and LLCC@DWCNT, respectively. ~\textbf{d},
The STEM image of a LLCC@DWCNT. ~\textbf{e}, The line profiles at
positions along the blue, red and green lines shown in (d),
represent empty DWCNT, LLCC@DWCNT and a thin most-inner
tube@DWCNT, respectively. ~\textbf{f}, Molecular model of
LLCC@DWCNT with (6,5) inner tube. Scale bar: 2 nm.}

\textsf{\textbf{Figure2: Calculation of the optimum diameter for
LLCC encapsulated in CNT.} DFT calculations about the interaction
energies between the LLCC and different chirality tubes: (4,4),
(5,5), (6,6) and (7,7). The inset: illustrations of different
CC@CNT structures. Notice that the distance between the LLCC and the wall of
the CNT remains similar for all host CNT. For the energy minimum, this distance is close to the radius of the CNT.}

\textsf{\textbf{Figure3: Raman spectra of LLCC@DWCNTs.} These
measurements were done using a 568.2~nm excitation wavelength.
~\textbf{a}, D-band, G-band and LLCC-bands of a sample annealed at
1460 $\,^{\circ}\mathrm{C}$ as compared to a pristine DWCNT. ~\textbf{b}, LLCC-band line shape analysis including four components. ~\textbf{c}, The area of the
LLCC-band as a function of annealing temperature for 30 min
annealing time. ~\textbf{d}, The area of the LLCC-band as a
function of annealing time at 1460 $\,^{\circ}\mathrm{C}$.
~\textbf{e}, RBM region for a sample annealed
at 1460 $\,^{\circ}\mathrm{C}$. ~\textbf{f}, Models of our most
abundant inner-tube-chiralites and chain
configurations.}

\textsf{\textbf{Figure 4 The temperature dependence of LLCC-band.}
~\textbf{a}, Raman spectra of LLCC@DWCNTs measured at different
temperatures. The spectra are are normalized to the G-band response. The inset shows the LLCC-band intensity in comparison to the G-line.
~\textbf{b}, FWHM of the individual components in LLCC-band as a
function of temperature. The solid line represents a model of a temperature dependent coupling two coupled modes, the dashed curve represents a fit to a model of a three-phonon decay. }

\textsf{\textbf{Figure 5: Comparison of the inverse lengths of
linear carbon chains with the Raman response}. Raman response of
polyynes as function of inverse length given by the number of
carbon atoms. The solid lines correspond to a linear fit. The top
line includes the data related to chains that are saturated or
suspended in a fluid medium \cite{Tabata06C,Wakabayashi07CPL,Agarwal13JoRS,Gibtner02CEJ}. The lower line contains
studies all embracing nanotubes taking into account an additional
downshift of 60 cm$^{-1}$ due to the interaction with the inner
tubes. The solid symbols are experimental results with confirmed
length \cite{Malard07PRB,Wakabayashi09EPJD,Nishide07JoPCC,Moura11PRB,Zhao11JoPCC} and the empty symbols correspond to the LLCC@DWCNTs
reported in this work and for LLCC@D/MWCNT \cite{Zhao11JoPCC,Zhao03PRL}. The partly
solid symbols are experimental results with length estimated by
TEM. The dashed line corresponds to a theoretical model using the
bond length alternation from Yang. et al.\cite{Yang07JoPCA}. }

\end{document}